\documentclass[prl,amsfonts,amssymb,floats,twocolumn,aps,superscriptaddress,showpacs]{revtex4}

\usepackage[dvips]{epsfig}
\usepackage{bm}

\begin{document} 
  
\title[]
{Variational cluster approach to correlated electron systems in low dimensions} 

\author{M. Potthoff}
\affiliation{
Institut f\"ur Theoretische Physik,
Universit\"at W\"urzburg,
Am Hubland,
97074 W\"urzburg,
Germany
}

\author{M. Aichhorn}
\affiliation{
Institut f\"ur Theoretische Physik,
Technische Universit\"at Graz,
Petersgasse 16,
8010 Graz,
Austria
}

\author{C. Dahnken}
\affiliation{
Institut f\"ur Theoretische Physik,
Universit\"at W\"urzburg,
Am Hubland,
97074 W\"urzburg,
Germany
}
 
\begin{abstract}
A self-energy-functional approach is applied to construct cluster approximations 
for correlated lattice models.
It turns out that the cluster-perturbation theory (S\'en\'echal et al, 
PRL {\bf 84}, 522 (2000)) and the cellular dynamical mean-field theory 
(Kotliar et al, PRL {\bf 87}, 186401 (2001)) are limiting cases of a more 
general cluster method.
Results for the one-dimensional Hubbard model are discussed with 
regard to boundary conditions, bath degrees of freedom and cluster size.
\end{abstract} 
 

\maketitle 

\begin{figure}[t]
\includegraphics[width=75mm]{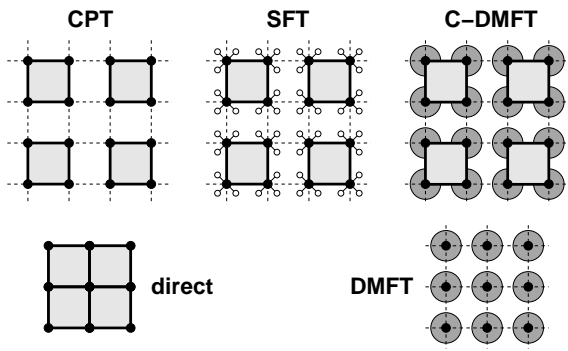}
\caption{
Reference systems considered within various cluster approximations for the 
$D=2$ Hubbard model.
Filled circles: on-site interaction.
Solid lines: intra-cluster hopping.
Dashed lines: inter-cluster hopping.
Open circles: additional $n_{\rm b}$ uncorrelated bath sites.
Big circles: bath with $n_{\rm b}=\infty$.
Common to all methods is (i) the numerical solution of the system of decoupled 
clusters and (ii) the subsequent coupling of the clusters via an RPA-like 
or Dyson equation. Bath parameters are determined self-consistently.
The self-energy-functional theory (SFT) comprises the extreme limits CPT 
($n_{\rm b}=0$) and C-DMFT ($n_{\rm b}=\infty$).
}
\label{fig:fig1}
\end{figure}

Low-dimensional systems of strongly interacting electrons, such as 
high-temperature superconductors, cuprate ladder compounds and organic 
conductors, currently form a focus of intense experimental and theoretical 
work.
It is generally accepted that many of the fascinating physical properties 
of these materials arise from different kinds of short-range spatial 
correlations as well as from different phases with long-range order.
On a low energy scale, this
can be studied within effective models such as the Hubbard
model \cite{hubbard}, for example.
As these models have to be considered in the intermediate- to strong-coupling
regime, weak-coupling perturbational approaches are inapplicable.
Cluster methods, which approximate the physics of the infinite system 
by solving the problem for a corresponding finite cluster, appear to be 
promising in this context as the interaction part can be treated numerically 
exact.
Currently, there are two different groups of cluster methods which are 
discussed intensively:

The point of reference for the first one is the {\em direct cluster approach}
\cite{Dag94}.
Using an exact-diagonalization or quantum Monte-Carlo technique, the effect
of short-range correlations can be studied by computing static and dynamic 
correlation functions for an isolated small cluster.
This direct method, however, suffers from the fact that phase transitions 
and long-range order cannot occur in a system of finite size. 
Furthermore, the spectral function consists of a comparatively small number
of poles.
This has caused the recent development of an extension called
{\em cluster-perturbation theory} (CPT) \cite{GV93,SPPL00,SPP02,ZEAH}.
The CPT procedure to calculate the one-electron Green's function 
${\bm G}$ \cite{matrix} is sketched in Fig.\ \ref{fig:fig1} for the Hubbard model:
(i) The lattice is divided into small clusters, and the inter-cluster 
hopping $\bm V$ is switched off.
The Green's function ${\bm G}'$ for the system of decoupled clusters 
(Hamiltonian $H'$) is calculated numerically.
(ii) The Green's function ${\bm G}$ of the lattice model $H$ is then 
approximated by
an RPA-like expression ${\bm G} = {\bm G}' ( 1 - {\bm V}{\bm G}')^{-1}$
\cite{matrix}.
It has been pointed out that this corresponds to the first order in a 
systematic expansion in the inter-cluster hopping \cite{SPPL00}.

The point of reference for the second type of cluster methods is the 
{\em dynamical mean-field theory} (DMFT) \cite{MV89,GKKR96}.
The lattice model $H$ is mapped onto an impurity model $H'$ consisting
of a correlated site coupled to an infinite number of uncorrelated 
``bath'' sites. 
The bath must be determined self-consistently.
As a mean-field theory, the DMFT directly works in the thermodynamic limit 
and is, thus, able to describe phases with long-range order.
Due to the locality of the self-energy \cite{MH89b}, however, it fails 
to incorporate the effects of short-range correlations.
This has been the reason for the development of cluster extensions of the 
DMFT \cite{SI95,HTZ+98,LK00,KSPB01}.
The main idea of the {\em cellular DMFT} (C-DMFT) \cite{KSPB01} is to replace 
the correlated impurity site by a finite cluster (see Fig.\ \ref{fig:fig1}) 
and to proceed as follows:
(i) The self-energy ${\bm \Sigma}$ is calculated numerically for the system 
of decoupled clusters with an uncorrelated bath attached to each of the 
correlated sites (Hamiltonian $H'$). 
(ii) The (approximate) Green's function of the lattice model $H$ is then 
obtained from ${\bm \Sigma}$ via the lattice Dyson equation:
${\bm G} = ( {\bm G}_0^{-1} - {\bm \Sigma})^{-1}$.
(iii) The parameters of the respective baths have to be recalculated from
${\bm G}$ and ${\bm \Sigma}$ via the C-DMFT selfconsistency condition
\cite{GKKR96,KSPB01}.
This requires to repeat the above steps until self-consistency is reached.

The first intention of this letter is to show that both the CPT and
the C-DMFT can be considered as extreme limits of a more general cluster
approach which is based on the self-energy-functional theory (SFT) proposed 
recently \cite{Pot03}.
This answers an open question \cite{SPP02} for the relation between the 
different cluster methods and unifies two approaches which appear 
to be rather different at first sight.
The presented cluster approach based on the SFT not only reproduces the 
CPT ($n_{\rm b}=0$) and the C-DMFT ($n_{\rm b}=\infty$) but also allows 
to construct approximations with an {\em arbitrary} number of bath sites 
$n_{\rm b}$. 
This intermediate approach does not waive any of the general merits such 
as causality and thermodynamical consistency.

Secondly, this letter shows that there is room for new conceptual ideas
beyond both the CPT and the C-DMFT:
(i) A consistent cluster approach can be constructed e.g.\ by attaching 
uncorrelated baths at the cluster boundary {\em only}.
(ii) Not only the bath parameters but also the on-site energies of and the 
hopping between the correlated sites may be determined in a self-consistent
(variational) way.
(iii) The question of boundary conditions can be answered by the method 
itself and need not be imposed by hand.
To discuss the relevance and advantages or disadvantages of these points,
some
numerical results will be presented for the one-dimensional ($D=1$) Hubbard 
model.

{\em Self-energy-functional theory (SFT):}
Consider a system of fermions on an infinite lattice with on-site Coulomb
interactions at temperature $T$ and chemical potential $\mu$.
Its Hamiltonian $H = H_0({\bm t}) + H_1({\bm U})$ consists of a 
one-particle part which depends on a set of hopping parameters ${\bm t}$ 
and an interaction part with Coulomb-interaction parameters ${\bm U}$.
The grand potential $\Omega$ can be obtained from the stationary point of
a self-energy functional
\begin{equation}
  \Omega_{\bm t}[{\bm \Sigma}] \equiv 
  {\rm Tr} \ln (- ({\bm G}_0^{-1} - {\bm \Sigma})^{-1}) + F[{\bm \Sigma}]
\label{eq:omega}
\end{equation}
as has been discussed in Ref.\ \cite{Pot03}. 
Here ${\bm G}_0 = 1 / (\omega + \mu - {\bm t})$ and $F[{\bm \Sigma}]$ is the 
Legendre transform of the Luttinger-Ward functional $\Phi[{\bm G}]$.
As the latter is constructed as an infinite series of renormalized 
skeleton diagrams \cite{LW60}, the self-energy functional is not known 
{\em explicitely}.
Nevertheless, the {\em exact} evaluation of $\Omega_{\bm t}[{\bm \Sigma}]$ 
and the determination of the stationary point is possible \cite{Pot03} on a 
{\em restricted} space ${\cal S}$ of trial self-energies 
${\bm \Sigma}({\bm t}') \in {\cal S}$.
Due to this restriction the procedure becomes approximative.

Generally, the space $\cal S$ consists of ${\bm t}'$ representable self-energies.
${\bm \Sigma}$ is termed ${\bm t}'$ representable if there are hopping 
parameters ${\bm t}'$ such that ${\bm \Sigma} = {\bm \Sigma}({\bm t}')$ is the 
exact self-energy of the model $H' = H_0({\bm t}') + H_1({\bm U})$ (``reference 
system'').
Note that both the original system $H$ and the reference system $H'$ must share 
the same interaction part.
For any ${\bm \Sigma}$ parameterized as ${\bm \Sigma}({\bm t}')$ 
we then have \cite{Pot03}:
\begin{equation}
   \Omega_{\bm t}[{\bm \Sigma}({\bm t}')] = \Omega' 
   + {\rm Tr} \ln (- ({\bm G}_0^{-1} - {\bm \Sigma}({\bm t}'))^{-1})
   - {\rm Tr} \ln (-{\bm G}') \: ,
\label{eq:om}
\end{equation}
where $\Omega'$, ${\bm G}'$, and ${\bm \Sigma}({\bm t}')$ are the grand 
potential, the Green's function and the self-energy of the reference system 
$H'$ while ${\bm G}_0$ is the free Green's function of $H$.
For a proper choice of ${\bm t}'$ (namely such that certain degrees of 
freedom in $H'$, e.g.\ those in different clusters, are decoupled), 
a (numerically) exact computation of these quantities is possible.
Hence, the self-energy functional (\ref{eq:om}) can be evaluated exactly 
for this ${\bm \Sigma} = {\bm \Sigma}({\bm t}')$.
A certain approximation is characterized by a choice for $\cal S$.
As ${\bm \Sigma}$ is parameterized by ${\bm t}'$, this means to specify a 
space of variational parameters ${\bm t}'$.
Any choice leads to a thermodynamically consistent approach
since, once the variational procedure is carried out, Eq.\ (\ref{eq:om}) provides 
an explicit expression for a thermodynamical potential.
For a further discussion of the general concept of the SFT see Ref.\ \cite{Pot03};
a detailed description of its practical application is given in Ref.\ \cite{Pot03a}.

{\em Cluster approximations:}
Fig.\ \ref{fig:fig1} illustrates the construction of cluster approximations
within the framework of the SFT.
To be specific, $H$ is taken to be the Hubbard model with nearest-neighbor
hopping.
Subdividing the infinite lattice into identical clusters of finite size,
$H'$ is obtained from $H$ by switching off the inter-cluster hopping and 
by switching on the hopping to new uncorrelated ($U_{\rm bath}=0$) bath 
sites (Fig.\ \ref{fig:fig1}, middle).
Both operations merely change the one-particle part of the Hamiltonian, i.e.\
${\bm t} \to {\bm t}'$, while the interaction part (${\bm U}$) remains fixed
-- as required.
To search for a stationary point on this space $\cal S$ of cluster-representable 
self-energies, one has to proceed as follows:
(i) Compute the self-energy ${\bm \Sigma}({\bm t}')$ of the reference system
for a given ${\bm t}'$.
(ii) Use Eq.\ (\ref{eq:om}) to evaluate $\Omega_{\bm t}[{\bm \Sigma}]$
at ${\bm \Sigma}={\bm \Sigma}({\bm t}')$.
(iii) Repeat steps (i) and (ii) for different ${\bm t}'$ to compute the function
$\Omega({\bm t}') \equiv \Omega_{\bm t}[{\bm \Sigma}({\bm t}')]$ and the 
stationary point ${\bm t}'_{\rm s}$ given by 
$\partial \Omega({\bm t}'_{\rm s}) /\partial {\bm t}'=0$.
As shown in Ref.\ \cite{Pot03a}, causality requirements are respected.

The variational adjustment of the intra-cluster one-particle parameters 
${\bm t}'$ can be looked upon as a (partial) compensation for the error 
introduced by the finite cluster size.
An inclusion of $n_{\rm b}$ bath sites per original correlated site enlarges 
the number of variational parameters and thereby the space ${\cal S}$. 
This is expected to (and does) improve the approximation (see results below).
In the limit of infinite cluster size (number of correlated sites within a 
cluster $N_c \to \infty$), the {\em exact} self-energy becomes ${\bm t}'$ representable
and therefore the cluster approximation itself becomes exact.
Since $N_c$ must be finite (small) in any practical calculation, one should 
focus on local quantities such as the on-site Green's 
function ${G}_{ii} = ({\bm G}_0^{-1} 
- {\bm \Sigma}({\bm t}'_{\rm s}))^{-1}_{ii}$, for example.

{\em CPT and C-DMFT:}
For a certain cluster approximation, it has to be specified which of the 
different intra-cluster one-particle parameters ${\bm t}'$ are treated as 
variational parameters.
The simplest idea is to consider the intra-cluster hopping as fixed at the 
original values, $t'_{ij}=t_{ij}$ (for $i,j$ in the same cluster), and not to 
switch on a hopping to bath sites (i.~e.\ not to introduce any bath sites).
In this case there is no variational parameter at all.
${\bm \Sigma}({\bm t}')$ is calculated once, and the Green's function ${\bm G}$ 
for the original model is obtained by the (lattice) Dyson equation
${\bm G}^{-1} = {\bm G}_0^{-1} - {\bm \Sigma}({\bm t}')$. 
As ${\bm \Sigma}({\bm t}')$ is the exact self-energy for $H'$ we also have:
${\bm \Sigma}({\bm t}') = {{\bm G}'_0}^{-1} - {{\bm G}'}^{-1}$ and consequently
${\bm G}^{-1} = {\bm G}_0^{-1} - {{\bm G}'_0}^{-1} + {{\bm G}'}^{-1} 
= {{\bm G}'}^{-1} - {\bm V}$ with ${\bm V}$ being the inter-cluster hopping.
As this is equivalent to the RPA-type equation mentioned above, one recovers
the CPT.

The C-DMFT is obtained by introducing a hopping $t'_{ir}$ to $n_{\rm b} =
\infty$ bath sites $r=1,...,n_{\rm b}$ per correlated site $i=1,...,N_{\rm c}$ 
and taking this hopping (``hybridization'') and the bath on-site energies 
$\epsilon_{i,r}'$ as variational parameters while for the correlated sites 
$t'_{ij}=t_{ij}$ is still fixed.
Assume that bath parameters $\{ t'_{ir} , \epsilon_{i,r}' \}$ can be found 
such that the C-DMFT self-consistency equation is fulfilled.
In Ref.\ \cite{KSPB01} this was given in ${\bm k}$-space representation 
(${\bm k}$ from the reduced Brillouin zone).
In the real-space representation the self-consistency equation reads
\begin{equation}
   ( 
   {\bm G}_0^{-1} - {\bm \Sigma}(\{ t'_{ir} , \epsilon_{i,r}' \}) 
   )_{ij}^{-1}
   = G'_{ij} \: ,
\label{eq:sc}
\end{equation}
where $i,j$ must belong to the same cluster.
This immediately implies that 
${\bm \Sigma}(\{ t'_{ir} , \epsilon_{i,r}' \})$ satisfies the SFT Euler 
equation $\partial \Omega({\bm t}') /\partial {\bm t}'=0$ or, calculating
the derivative,
\begin{equation}
   T \sum_{\omega} \sum_{ij}
   \left( 
   \frac{1}{{\bm G}_0^{-1} - {\bm \Sigma}({\bm t}')} 
   -  {\bm G}' \right)_{ji} 
   \frac{\partial \Sigma_{ij}({\bm t}')}
        {\partial {{\bm t}'}}
   = 0 \: .
\nonumber \\   
\label{eq:euler}
\end{equation}
This holds since the ``projector'' 
$\partial \Sigma_{ij}({\bm t}') / \partial {{\bm t}'} = 0$ if $i,j$ 
belong to different clusters as these are decoupled in the reference
system.
We conclude that the self-energy functional is stationary at the C-DMFT
self-energy.

In principle, approximations may also be constructed in reciprocal $\bm k$ space.
For the Hubbard model, however, there is no simple reference system as the 
interaction part is non-local in $\bm k$ space.
In particular, it is not possible to recover the dynamical cluster approximation 
(DCA) \cite{HTZ+98} within the SFT.

{\em ``Intermediate'' approach:}
The C-DMFT self-consistency equation can generally be fulfilled 
for $n_{\rm b} = \infty$ only.
Within a cluster approach based on the SFT there are no formal problems, 
however, if $n_{\rm b} < \infty$. 
A finite $n_{\rm b}$ yields an approximation {\em inferior} as compared to 
$n_{\rm b} = \infty$ (C-DMFT) and {\em superior} as compared to $n_{\rm b} = 0$ 
(CPT) as there are less or more variational parameters, respectively.
For $0 < n_{\rm b} < \infty$ the parameters must be found to satisfy
the Eq.\ (\ref{eq:euler}).

Since the convergence with respect to $n_{\rm b}$ appears to be rapid for local 
physical quantities \cite{Pot03} and since the cluster Hilbert-space dimension 
increases exponentially with $n_{\rm b}$, approximations with small $n_{\rm b}$ 
(or even $n_{\rm b}=0$, CPT) appear advantageous.
Anyway, the bath concept must become irrelevant in the limit $N_{\rm c} \to \infty$.
On the other hand, there are good reasons to introduce bath sites:
Depending on the dimensionality of the problem, it can be the 
description of the local (temporal) correlations that needs to be improved
in first place.
Note that for $N_{\rm c}=1$ and $n_{\rm b}= \infty$ the DMFT is recovered
which represents the {\em exact} solution for $D=\infty$ \cite{GKKR96}.
Furthermore, bath sites can serve as particle reservoirs which will be 
essential for a proper description of filling dependencies.
Finally, the presence of bath sites may also facilitate practical calculations
to treat the reference system, e.g.\ by an attenuation of the sign problem in 
the context of a Hirsch-Fye like QMC approach \cite{HF86}.

{\em Intra-cluster hopping determined variationally:}
For both the CPT and the C-DMFT the hopping between correlated sites is 
fixed at $t'_{ij}=t_{ij}$ (for the C-DMFT, this is even a necessary 
condition to satisfy Eq.\ (\ref{eq:sc}) as can easily be seen from a 
high-frequency expansion).
Contrary, within the SFT there is {\em a priori} no reason to fix $t'_{ij}$.
A cluster approximation with $t'_{ij}$ determined variationally represents
another ``intermediate'' approach as shall be discussed in the following.

\begin{figure}[t]
\includegraphics[width=80mm]{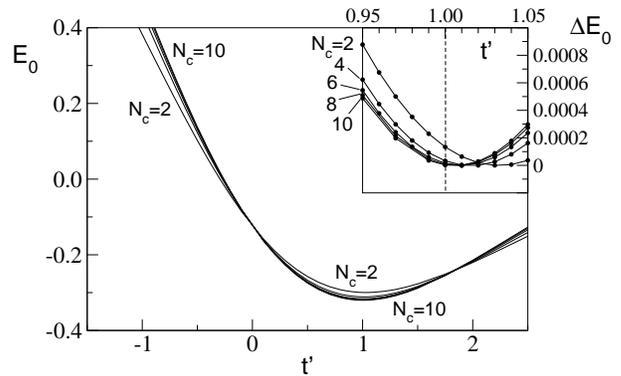}
\caption{
$E_0 \equiv \Omega + \mu \langle N \rangle$ vs.\ $t'$ as obtained by evaluating
the self-energy functional $\Omega = \Omega[{\bm \Sigma}(t')]$.
Original system $H$: $D=1$ Hubbard model with n.n.\ hopping $t=1$ and 
$U=8$ for $T=0$ and $\mu=U/2$ (half-filling).
Reference system $H'$: set of decoupled Hubbard chains with $N_{\rm c}$ sites 
each; $N_{\rm c}=2,4,6,8,10$ as indicated.
Variational parameter: n.n.\ hopping $t'$ of $H'$.
The inset shows $\Delta E_0 = E_0 - E_{0,{\rm min}}$ vs.\ $t'$.
}
\label{fig:fig2}
\end{figure}

Numerical calculations have been performed for the $D=1$ Hubbard model. 
Instead of solving the Euler equation (\ref{eq:euler}), we have directly 
evaluated the self-energy functional according to Eq.\ (\ref{eq:om}).
The reference system $H'$ is taken to be a set of decoupled Hubbard chains 
with $N_{\rm c}$ correlated sites each.
For $N_{\rm c} \le 10$ (no additional bath sites), the ground-state energy 
$E_0'$ (and thereby $\Omega' = E_0' - \mu \langle N \rangle'$ for $T=0$) and 
the Green's function ${\bm G}'$ are computed using the standard Lanczos algorithm
\cite{LG93}.
The self-energy is obtained as 
${\bm \Sigma}({\bm t}') = {{\bm G}'_0}^{-1} - {{\bm G}'}^{-1}$.
In Eq.\ (\ref{eq:om}), the trace ``Tr'' consists of a sum over a complete set 
of one-particle quantum numbers and (after analytical continuation to the 
real $\omega$ axis) a frequency integration.
To keep the calculations simple, only a single variational parameter is 
taken into account.

Fig.\ \ref{fig:fig2} shows $E_0(t') = \Omega(t') + \mu \langle N \rangle'$ 
with $\Omega(t') \equiv \Omega_{\bm t}[{\bm \Sigma}(t')]$ where $t'$ is the 
nearest-neighbor hopping within the cluster. 
Clearly, $E_0$ is stationary for $t'$ very close to $t$ but $t'\ne t$.
The effect is most obvious for the smallest cluster size $N_{\rm c}=2$ 
(see inset).
We conclude that a variational determination of the hopping between the 
correlated sites in fact improves the approximation.
Surprisingly, however, the improvement is almost negligible for reasonable 
$N_{\rm c}$.
Similar results are obtained when different selected hopping parameters are 
varied.

\begin{figure}[t]
\centerline{\includegraphics[width=80mm]{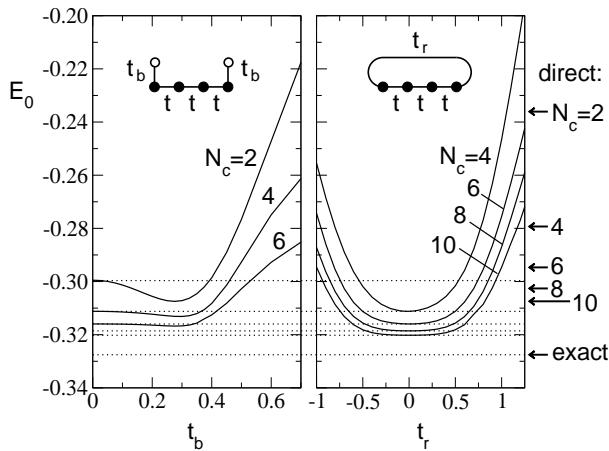}}
\caption{
$E_0$ as in Fig.\ \ref{fig:fig2} but vs.\ different variational parameters:
n.n.\ hopping $t_{\rm b}$ to uncorrelated bath sites (left),
hopping $t_r$ between the edge sites (right).
Dashed lines: $E_0$ without parameter optimization.
Arrows: $E_0$ calculated for an isolated cluster.
Exact result for $E_0$ from Ref.\ \cite{LW68}.
}
\label{fig:fig3}
\end{figure}

{\em Different variational parameters:}
It is possible to construct a consistent cluster approach by attaching 
uncorrelated bath sites only at the boundaries of the respective cluster.
Compared to the C-DMFT, this choice represents a certain restriction of the 
parameter set which is motivated by the expectation that here bath degrees 
of freedom compensate for the finite-size errors most efficiently.
Fig.\ \ref{fig:fig3} (left) shows that switching on the hopping $t_{\rm b}$ 
to two bath sites in fact lowers the minimal $E_0$.
With increasing $N_{\rm c}$ the energy difference 
$E_0(t_{\rm b}=t_{\rm b}^{\rm min}) - E_0(0)$ 
decreases, and $E_0(t_{\rm b})$ becomes almost flat for small $t_{\rm b}$,
as expected.
Note that the binding energy gain due to inclusion of two bath sites
$|E_0(N_{\rm c},t_{\rm b}^{\rm min}) - E_0(N_{\rm c},0)|$ is always smaller 
than the gain $|E_0(N_{\rm c}+2,0) - E_0(N_{\rm c},0)|$ due to a larger 
cluster: Introducing bath sites is less efficient as increasing the
cluster size, at least for $D=1$.
Interestingly, the convergence of $E_0$ with increasing $N_{\rm c}$ appears to
be much faster as compared to the direct cluster method 
(see arrows in Fig.\ \ref{fig:fig3}).

Within the CPT the question of the correct boundary conditions is decided
{\em a posteriori} by inspection of the respective results for the spectral
density \cite{SPP02}.
Here we introduce a hopping parameter $t_{\rm r}$ between the edge 
sites and let the method ``decide'' by itself.
As is seen in Fig.\ \ref{fig:fig3} (right), a minimum for $E_0$ is obtained
at $t_{\rm r} = 0$ (open boundary conditions) while there is no indication
for a stationary point at $t_{\rm r} = t$ (periodic boundary conditions).

{\em Conclusion:}
The self-energy-functional theory has been shown to provide a consistent
and very general framework for the construction of variational cluster 
approximations including the CPT and the C-DMFT.
The extreme flexibility with regard to the choice of variational parameters
offers a variety of further methodical developments and applications.

\begin{acknowledgments}
It is a pleasure to thank E.\ Arrigoni and W.\ Hanke for discussions. 
M.\ Aichhorn is supported by DOC [Doctoral Scholarship Program of the 
Austrian Academy of Sciences] and acknowledges hospitality at the 
physics institute in W\"urzburg. 
The work is supported by the DFG (SFB~290) and KONWIHR (CUHE).
\end{acknowledgments}


\begin{thebibliography}{19}
\expandafter\ifx\csname natexlab\endcsname\relax\def\natexlab#1{#1}\fi
\expandafter\ifx\csname bibnamefont\endcsname\relax
  \def\bibnamefont#1{#1}\fi
\expandafter\ifx\csname bibfnamefont\endcsname\relax
  \def\bibfnamefont#1{#1}\fi
\expandafter\ifx\csname citenamefont\endcsname\relax
  \def\citenamefont#1{#1}\fi
\expandafter\ifx\csname url\endcsname\relax
  \def\url#1{\texttt{#1}}\fi
\expandafter\ifx\csname urlprefix\endcsname\relax\def\urlprefix{URL }\fi
\providecommand{\bibinfo}[2]{#2}
\providecommand{\eprint}[2][]{\url{#2}}

\bibitem[{\citenamefont{Hubbard}((1963))}]{hubbard}
\bibinfo{author}{\bibfnamefont{J.}~\bibnamefont{Hubbard}},
  \bibinfo{journal}{Proc. R. Soc. London A} \textbf{\bibinfo{volume}{276}},
  \bibinfo{pages}{238} (\bibinfo{year}{1963});
\bibinfo{author}{\bibfnamefont{M.~C.} \bibnamefont{Gutzwiller}},
  \bibinfo{journal}{Phys. Rev. Lett.} \textbf{\bibinfo{volume}{10}},
  \bibinfo{pages}{159} (\bibinfo{year}{1963});
\bibinfo{author}{\bibfnamefont{J.}~\bibnamefont{Kanamori}},
  \bibinfo{journal}{Prog. Theor. Phys. (Kyoto)} \textbf{\bibinfo{volume}{30}},
  \bibinfo{pages}{275} (\bibinfo{year}{1963}).

\bibitem[{\citenamefont{Dagotto}((1994))}]{Dag94}
\bibinfo{author}{\bibfnamefont{E.}~\bibnamefont{Dagotto}},
  \bibinfo{journal}{Rev. Mod. Phys.} \textbf{\bibinfo{volume}{66}},
  \bibinfo{pages}{763} (\bibinfo{year}{1994}).

\bibitem[{\citenamefont{Gros and Valenti}((1993))}]{GV93}
\bibinfo{author}{\bibfnamefont{C.}~\bibnamefont{Gros}} \bibnamefont{and}
  \bibinfo{author}{\bibfnamefont{R.}~\bibnamefont{Valenti}},
  \bibinfo{journal}{Phys. Rev. B} \textbf{\bibinfo{volume}{48}},
  \bibinfo{pages}{418} (\bibinfo{year}{1993}).

\bibitem[{\citenamefont{S\'en\'echal et~al.}(2000)\citenamefont{S\'en\'echal,
  P\'erez, and Pioro-Ladri\`ere}}]{SPPL00}
\bibinfo{author}{\bibfnamefont{D.}~\bibnamefont{S\'en\'echal et al}},
  \bibinfo{journal}{Phys. Rev. Lett.} \textbf{\bibinfo{volume}{84}},
  \bibinfo{pages}{522} (\bibinfo{year}{2000}).
  
\bibitem[{\citenamefont{S\'en\'echal et~al.}(2002)\citenamefont{S\'en\'echal,
  P\'erez, and Plouffe}}]{SPP02}
\bibinfo{author}{\bibfnamefont{D.}~\bibnamefont{S\'en\'echal et al}},
  \bibinfo{journal}{Phys. Rev. B} \textbf{\bibinfo{volume}{66}},
  \bibinfo{pages}{075129} (\bibinfo{year}{2002}).
  
\bibitem[{\citenamefont{Zacher et~al.}((2000))\citenamefont{Zacher, Eder,
  Arrigoni, and Hanke}}]{ZEAH}
\bibinfo{author}{\bibfnamefont{M.~G.} \bibnamefont{Zacher et al}},
  \bibinfo{journal}{Phys. Rev. Lett.} \textbf{\bibinfo{volume}{85}},
  \bibinfo{pages}{2585} (\bibinfo{year}{2000});
  \bibinfo{journal}{Phys. Rev. B} \textbf{\bibinfo{volume}{65}},
  \bibinfo{pages}{045109} (\bibinfo{year}{2002}).
    
\bibitem{matrix}
A matrix notation is used: Consider e.g.\ the Hubbard model.
Then ${\bm G}$ and $\bm V$, for example, have the elements 
$G_{ij}(\omega)$ and $V_{ij}$ where $i,j$ refer to sites. 
Furthermore, $\mbox{Tr} \, {\bm A} = 2 T \sum_{\omega,i} 
A_{ii}(i\omega)$ for a function ${\bm A}$ of the imaginary 
Matsubara frequencies $i\omega = i(2n+1) \pi T$ with integer $n$.
The spin index is suppressed.

\bibitem[{\citenamefont{Metzner and Vollhardt}((1989))}]{MV89}
\bibinfo{author}{\bibfnamefont{W.}~\bibnamefont{Metzner}} \bibnamefont{and}
  \bibinfo{author}{\bibfnamefont{D.}~\bibnamefont{Vollhardt}},
  \bibinfo{journal}{Phys. Rev. Lett.} \textbf{\bibinfo{volume}{62}},
  \bibinfo{pages}{324} (\bibinfo{year}{1989}).

\bibitem[{\citenamefont{Georges et~al.}((1996))\citenamefont{Georges, Kotliar,
  Krauth, and Rozenberg}}]{GKKR96}
\bibinfo{author}{\bibfnamefont{A.}~\bibnamefont{Georges et al}},
  \bibinfo{journal}{Rev. Mod. Phys.} \textbf{\bibinfo{volume}{68}},
  \bibinfo{pages}{13} (\bibinfo{year}{1996}).

\bibitem[{\citenamefont{M\"uller-Hartmann}((1989))}]{MH89b}
\bibinfo{author}{\bibfnamefont{E.}~\bibnamefont{M\"uller-Hartmann}},
  \bibinfo{journal}{Z. Phys. B} \textbf{\bibinfo{volume}{74}},
  \bibinfo{pages}{507} (\bibinfo{year}{1989}).

\bibitem[{\citenamefont{Schiller and Ingersent}((1995))}]{SI95}
\bibinfo{author}{\bibfnamefont{A.}~\bibnamefont{Schiller}} \bibnamefont{and}
  \bibinfo{author}{\bibfnamefont{K.}~\bibnamefont{Ingersent}},
  \bibinfo{journal}{Phys. Rev. Lett.} \textbf{\bibinfo{volume}{75}},
  \bibinfo{pages}{113} (\bibinfo{year}{1995}).

\bibitem[{\citenamefont{Hettler et~al.}((1998))\citenamefont{Hettler,
  Tahvildar-Zadeh, Jarrell, Pruschke, and Krishnamurthy}}]{HTZ+98}
\bibinfo{author}{\bibfnamefont{M.~H.} \bibnamefont{Hettler et al}},
  \bibinfo{journal}{Phys. Rev. B} \textbf{\bibinfo{volume}{58}},
  \bibinfo{pages}{R7475} (\bibinfo{year}{1998}).

\bibitem[{\citenamefont{Lichtenstein and Katsnelson}((2000))}]{LK00}
\bibinfo{author}{\bibfnamefont{A.~I.} \bibnamefont{Lichtenstein}}
  \bibnamefont{and} \bibinfo{author}{\bibfnamefont{M.~I.}
  \bibnamefont{Katsnelson}}, \bibinfo{journal}{Phys. Rev. B}
  \textbf{\bibinfo{volume}{62}}, \bibinfo{pages}{R9283}
  (\bibinfo{year}{2000}).

\bibitem[{\citenamefont{Kotliar et~al.}((2001))\citenamefont{Kotliar, Savrasov,
  P\'alsson, and Biroli}}]{KSPB01}
\bibinfo{author}{\bibfnamefont{G.}~\bibnamefont{Kotliar et al}},
  \bibinfo{journal}{Phys. Rev. Lett.} \textbf{\bibinfo{volume}{87}},
  \bibinfo{pages}{186401} (\bibinfo{year}{2001}).

\bibitem[{\citenamefont{Potthoff}(2003)}]{Pot03}
\bibinfo{author}{\bibfnamefont{M.}~\bibnamefont{Potthoff}},
  \bibinfo{journal}{Eur. Phys. J. B} \textbf{\bibinfo{volume}{32}},
  \bibinfo{pages}{429} (\bibinfo{year}{2003}).

\bibitem[{\citenamefont{Luttinger and Ward}((1960))}]{LW60}
\bibinfo{author}{\bibfnamefont{J.~M.} \bibnamefont{Luttinger}}
  \bibnamefont{and} \bibinfo{author}{\bibfnamefont{J.~C.} \bibnamefont{Ward}},
  \bibinfo{journal}{Phys. Rev.} \textbf{\bibinfo{volume}{118}},
  \bibinfo{pages}{1417} (\bibinfo{year}{1960}).

\bibitem[{\citenamefont{Potthoff}(2003)}]{Pot03a}
\bibinfo{author}{\bibfnamefont{M.}~\bibnamefont{Potthoff}},
  \bibinfo{journal}{preprint cond-mat/0306278}.

\bibitem[{\citenamefont{Hirsch and Fye}((1986))}]{HF86}
\bibinfo{author}{\bibfnamefont{J.~E.} \bibnamefont{Hirsch}} \bibnamefont{and}
  \bibinfo{author}{\bibfnamefont{R.~M.} \bibnamefont{Fye}},
  \bibinfo{journal}{Phys. Rev. Lett.} \textbf{\bibinfo{volume}{56}},
  \bibinfo{pages}{2521} (\bibinfo{year}{1986}).

\bibitem[{\citenamefont{Lin and Gubernatis}((1993))}]{LG93}
\bibinfo{author}{\bibfnamefont{H.~Q.} \bibnamefont{Lin}} \bibnamefont{and}
  \bibinfo{author}{\bibfnamefont{J.~E.} \bibnamefont{Gubernatis}},
  \bibinfo{journal}{Comput. Phys.} \textbf{\bibinfo{volume}{7}},
  \bibinfo{pages}{400} (\bibinfo{year}{1993}).

\bibitem[{\citenamefont{Lieb and Wu}((1968))}]{LW68}
\bibinfo{author}{\bibfnamefont{E.~H.} \bibnamefont{Lieb}} \bibnamefont{and}
  \bibinfo{author}{\bibfnamefont{F.~Y.} \bibnamefont{Wu}},
  \bibinfo{journal}{Phys. Rev. Lett.} \textbf{\bibinfo{volume}{20}},
  \bibinfo{pages}{1445} (\bibinfo{year}{1968}).
    
\end{thebibliography}
\end{document}